  \documentclass[twocolumn]{aa} % for a paper on 1 column
\usepackage{booktabs}

\usepackage{graphicx}
\usepackage{txfonts}
\newcommand{\hbop}{Hale-Bopp}
\newcommand{\hya}{Hyakutake}

\newcommand{\grat}{G/R ratio}
\newcommand{\chris}{C/2006~W3 (Christensen)}
\newcommand{\chrisw}{C/2006~W3}
\newcommand{\linear}{C/2003~K4 (LINEAR)}
\newcommand{\linek}{C/2003~K4}
\newcommand{\neat}{C/2001~Q4 (NEAT)}
\newcommand{\neatq}{C/2001~Q4}
\newcommand{\garad}{C/2009~P1 (Garradd)}
\newcommand{\garadp}{C/2009~P1}
\newcommand{\sid}{C/2007~Q3 (Siding Spring)}
\newcommand{\sidq}{C/2007~Q3}
\newcommand{\wild}{116P/Wild~4}
\newcommand{\wildp}{116P}
\newcommand{\old}{O($^1$D)}
\newcommand{\ols}{O($^1$S)}
\usepackage{hyperref}
\hypersetup{
bookmarks=false,        % show bookmarks bar?
unicode=false,          % non-Latin characters in Acrobat’s bookmarks
pdftoolbar=true,        % show Acrobat’s toolbar?
pdfmenubar=true,        % show Acrobat’s menu?
pdffitwindow=false,     % window fit to page when opened
pdfstartview={FitH},    % fits the width of the page to the window
pdftitle={ },    % title
pdfauthor={Susarla Raghuram},   % author
pdfsubject={ },   % subject of the document
pdfkeywords={ },     % list of keywords
pdfnewwindow=true,      % links in new window
colorlinks=true,        % false: boxed links; true: colored links
linkcolor=black,         % color of internal links
citecolor=black,        % color of links to bibliography
filecolor=magenta,      % color of file links
urlcolor=red            % color of external links
}
\begin{document}

   \title{Photochemistry of  atomic oxygen  green and red-doublet emissions in 
comets at larger heliocentric distances}
%    \subtitle{I. Overviewing the $\kappa$-mechanism}
   \author{ Susarla Raghuram\thanks{Corresponding author : raghuramsusarla@gmail.com}  %\inst{1}%           
          \ \and  
         \ Anil Bhardwaj%\inst{1}
          }
   \institute{Space Physics Laboratory, Vikram Sarabhai Space Center,  Trivandrum
695022, India}. 
              \\
%               \email{ }
%          \and
%              University of Alexandria, Department of Geography, ...\\
%              \email{raghuramsusarla@gmail.com}
%              \thanks{The university of heaven temporarily does not
%                      accept e-mails}

     \date{\centering \bf\Large{Accepted for publication}}

% \abstract{}{}{}{}{} 
% 5 {} token are mandatory
 
  \abstract
  % context heading (optional)
  % {} leave it empty if necessary  
   {In comets the atomic oxygen  green (5577 \AA) to red-doublet (6300, 6364 \AA)
   emission intensity ratio (\grat)
 of 0.1 has been used to confirm
H$_2$O as the parent species producing forbidden oxygen emission lines. The larger 
($>$0.1) value of \grat\  observed in a few comets is ascribed to the presence of
higher CO$_2$ and CO relative abundances in the cometary coma.}
  % aims heading (mandatory)
   {We aim to study the effect of CO$_2$ and CO relative abundances on the observed \grat\
in comets observed at large ($>$2 au) heliocentric distances by accounting for
important
 production and loss processes of \ols\ and \old\ atoms in the cometary coma.}
  % methods heading (mandatory)
   {Recently we have developed a coupled chemistry-emission model to study 
photochemistry of  \ols\ and \old\ atoms and the production of green and red-doublet
emissions in  comets \hya\ and \hbop. In the present work we applied the model
to six comets where green and red-doublet emissions are observed when they are beyond
2 au 
from the Sun.}
  % results heading (mandatory)
   {The collisional quenching of \ols\ and \old\ can alter the \grat\ more 
significantly than that due to change in the relative abundances of CO$_2$ and CO. In
a water-dominated cometary coma and with significant ($>$10\%) CO$_2$ relative
abundance, photodissociation of H$_2$O mainly governs the red-doublet
 emission, whereas  CO$_2$ controls the green line emission.  If a comet has equal
composition of CO$_2$ and H$_2$O, then $\sim$50\% of red-doublet emission intensity
is controlled by the photodissociation of CO$_2$. The role of CO photodissociation
is insignificant in producing both green and red-doublet emission lines and
consequently in determining the \grat. 
Involvement of multiple production sources in the \ols\ formation may be the reason for 
the observed higher green line width than that of  red lines.
The \grat\ values and green and red-doublet line widths calculated by the
model are consistent with the observation.}
  % conclusions heading (optional), leave it empty if necessary 
   {Our model calculations suggest that in  low gas production
 rate comets the \grat\ greater than 0.1 can be used to constrain the upper
limit of CO$_2$ relative abundance
provided the slit-projected area on the coma is larger than the collisional zone.
If a comet has equal abundances of CO$_2$ and H$_2$O,
then the red-doublet emission is significantly ($\sim$50\%) controlled by CO$_2$ 
photodissociation
and thus the \grat\ is not suitable for estimating  CO$_2$ relative abundance.}


   \maketitle
%  \titlerunning[1]{photochemistry of green and red-doublet emissions.}    
% \authorruning[1]{Bhradwaj and Raghuram}
%
% ________________________________________________________________

\section{Introduction}
 Green (5577 \AA) and red-doublet (6300, 6364 \AA) emissions are due to the 
electronic transition of oxygen atoms from metastable $^1$S and $^1$D states,
respectively, to the
ground $^3$P state.
Since \ols\ and \old\ are metastable states, resonance fluorescence  by solar photons 
 is not an effective excitation mechanism for populating these states.
Dissociative excitation of O-bearing neutrals by photons and photoelectrons, 
and thermal recombination of atomic oxygen constituted ions  are the sources
 of  these metastable states in the cometary coma \citep{Bhardwaj02,Bhardwaj12,Raghuram13}.
 Most of the comets observed around heliocentric distance
of 1 au have  H$_2$O as the principal constituent in the cometary coma
\citep{Bockelee04}.
Based on the theoretical work of 
\cite{Festou81} green to red-doublet emission intensity
 ratio (hereafter \grat) of 0.1 has customarily been
used as a benchmark to confirm the parent source of these prompt emissions
 as H$_2$O in several comets observed around 1 au from the Sun \citep{Cochran84,
 Cochran08, Morrison97, Zhang01,Cochran01, Furusho06,Capria05,Capria08,Capria10}. The
observed \grat\ of more than 0.1 has been attributed to higher relative abundances of
CO$_2$ and CO \citep{Furusho06,Capria10,McKay12,Decock13}.
Since no experimental cross section or yield for the production of \ols\
from H$_2$O is available in the literature, the calculated photorates of
\cite{Festou81} have been questioned by \cite{Huestis06}.  In a H$_2$O-dominated
cometary coma, the red-doublet emission intensity is determined by formation and 
destruction rates of \old\ \citep{Bhardwaj02,Bhardwaj12,Raghuram13}.
Since the red-doublet emission is mainly 
governed by photodissociation of H$_2$O,  the observed intensity of 6300 \AA\ 
has been used to estimate the production rate as well as to understand the spatial 
distribution of H$_2$O in the cometary coma
\citep[e.g.][]{Delsemme76,Delsemme79,Fink84,Schultz92,
Morgenthaler01,Furusho06}.

We  have recently developed a coupled chemistry-emission model for the production of
 green and red-doublet emissions by accounting for important production and loss
mechanisms of  \ols\ and \old\ atoms. The model has been applied to comets \hya\  
\citep{Bhardwaj12} and \hbop\
 \citep{Raghuram13}. Our model
calculations showed that in a H$_2$O-dominated cometary coma more than 90\%
of the \old\ is populated via photodissociative excitation of H$_2$O and the rest 
through
photodissociation of CO$_2$ and CO. We also demonstrated that the \grat\ 
depends not only on the photochemistry involved in the formation of \old\ and \ols,
but also on the projected area observed on the comet, which is a function of slit
dimension and geocentric distance of the comet \citep{Bhardwaj12}. The model
calculations on comets \hya\ and \hbop\ showed that the intensity of the [OI] 6300
\AA\ line is largely governed by photodissociation of H$_2$O, whereas the [OI] 5577
\AA\
 emission line is mainly controlled by the photodissociation of both H$_2$O and
CO$_2$.  It is also suggested that  CO$_2$ can produce \ols\ more efficiently than
H$_2$O. The calculated mean excess energy profiles in  various photodissociation
processes showed that the photodissociation of CO$_2$ can produce \ols\ with higher
excess velocity compared to the photodissociation of H$_2$O \citep{Raghuram13}.
All these calculations are carried out at $\sim$1 au.

 At larger heliocentric distances 
the cometary coma  is composed of larger proportions of CO and CO$_2$
than at 1 au \citep{Meech04,Crovisier99,Biver97,Biver99,Bockelee04,Bockelee10}.
At heliocentric distances of more than 2 au the prompt emissions of atomic oxygen  
are observed in several comets,
 viz.  \sid, \chris, \garad, \neat, \wild, and \linear\
\citep{Furusho06,McKay12,Decock13}.
Assuming that CO$_2$ and CO are the main sources of green and red-doublet
emissions, the observed \grat\ in comets at large heliocentric distances ($>$ 2 au) 
has been used to estimate the CO$_2$ abundance in comets \citep{Decock13,McKay12}.

The present study is aimed at studying the photochemistry of \ols\ and \old\ atoms and
associated green and red-doublet emission production mechanisms in the above mentioned
six comets at larger heliocentric distance ($>$ 2 au) where the gas production rate
of CO can be equal to that of H$_2$O.
One of the objectives of the study is to verify whether the \grat\ value can be used
to infer the CO$_2$ relative abundance, with respect to H$_2$O, in the comets that 
are observed at larger heliocentric distances. In this study we have shown  that even
at large heliocentric distances, the photodissociation of CO is only a minor
source of \ols\ and \old\ atoms, and its impact on the \grat\ is negligible.
The red-doublet emission intensity is mainly governed by H$_2$O, while the
green line emission intensity is controlled by CO$_2$.
 We have also demonstrated that collisional quenching can significantly change the
observed \grat\ and that its impact on the \grat\ is much greater than that due to
variation in the CO$_2$ and H$_2$O abundances.

\section{Model}
The details of model and the photochemical reactions considered
in the  model 
 are presented in our previous works \citep{Bhardwaj12,Raghuram13}.
Here we present the input parameters that have been used in the
model for the calculation of green and red-doublet emission intensities for the 
observed conditions of six comets (viz. \wild, \linear, \sid, \chris, \garad,
\neat).  The photochemical reaction network and
cross sections of photon and electron impact processes that have been considered in our
previous work remain the same for the present calculation. 
The photoelectron impact excitation reactions are accounted for by degrading solar 
extreme ultraviolet (EUV) generated  photoelectrons and electron impact cross
sections in the cometary coma using the  analytical yield spectrum (AYS) technique
which is based on the Monte Carlo method. Details of the AYS approach and the method
for calculating
photoelectron flux and excitation rates are given in our earlier papers and
references  therein \citep{Bhardwaj99a,Haider05,Bhardwaj11,Raghuram12,Bhardwaj12}.
 The production and loss mechanisms for the \ols\ and \old\ considered in the model
calculations are presented in our previous papers \citep{Bhardwaj12,Raghuram13}.
 Only the dominant O-bearing neutral species H$_2$O, CO$_2$, and CO are considered in
 the present  model.

The neutral gas production rates used in the model calculations  for different comets
during observation periods of oxygen emission lines  are tabulated in 
Table~\ref{tab-comet}. 
In some comets these gas production rates are  not measured, and so we have made a
reasonable 
approximation to incorporate CO$_2$ and CO in the model. However, we vary the CO$_2$
and CO relative abundances on these comets to assess the impact on the green and red-doublet
emission intensities and subsequently on the \grat.

\begin{table*}[tbh]
\renewcommand{\thefootnote}{\fnsymbol{footnote}}
\begin{center}
 \caption{Observational conditions (slit dimension, heliocentric ($r$),  and
geocentric ($\Delta$) distances)
of various comets, corresponding H$_2$O production rates and CO$_2$ and 
CO relative abundances relative to H$_2$O, and
 comparison of calculated green to red-doublet emission intensity ratios (\grat)  
with the observations.}
\scalebox{0.88}[1]{
\label{tab-comet}
\begin{tabular}{@{}lllllllllllllllll@{}}
\toprule
  Comet & $r$ & $\Delta$ & Slit dimension & Q(H$_2$O) & CO$_2$\footnotemark[5] & 
CO\footnotemark[5] &
\multicolumn{2}{l} {\grat} &Reference\footnotemark[9]\\ 
   & (au) & (au) & ($''$ $\times$ $''$) & (s$^{-1}$) & (\%) & (\%) &   cal. &obs.
&\\ 
\midrule
  \wild & 2.40 & 1.4  & 8 $\times$ 1     & 1 $\times$ 10$^{27}$\footnotemark[1] & 10 
& 20& 0.09 &
  0.15 &  \cite{Furusho06}\\  
  \linear & 2.60 & 2.36 &  0.80 $\times$ 11 & 1 $\times$ 10$^{29}$\footnotemark[5] &
10\footnotemark[6] & 25\footnotemark[6] & 0.09 &
0.09  & \cite{Decock13}\\  
  \sid  & 2.96 & 2.48 & 3.20 $\times$ 1.6   & 4 $\times$ 10$^{27}$\footnotemark[1]
& 17& 10& 0.12 & 0.20&  \cite{McKay12}\\  
  \chris & 3.13 & 2.35 & 3.20 $\times$ 1.6   & 2.0 $\times$ 10$^{28}$\footnotemark[1]
& 42& 98& 0.18 & 0.24 $\pm$ 0.08 & \cite{McKay12}\\ 
  \garad & 3.25  & 3.50  & 0.44 $\times$ 12 & 2.3 $\times$
10$^{27}$\footnotemark[3]   & 25\footnotemark[6] & 100\footnotemark[6]  & 0.14 & 0.21
&   \cite{Decock13}\\  
  \neat & 3.70 & 3.40 & 0.45 $\times$ 11  & 3.8 $\times$ 10$^{27}$\footnotemark[2] &
75\footnotemark[6]& 100\footnotemark[6] & 0.23 & 0.33 & \cite{Decock13}\\  
 comet X\footnotemark[7] & 3.70 & 3.40 & 0.45 $\times$ 11  & 4 $\times$ 10$^{27}$ &
100 & 100 & 0.25 & -- & -- \\  
\bottomrule
\end{tabular}}
\end{center}
\footnotemark[1]{\cite{Ootsubo12};}
\footnotemark[2]{\cite{Combi09};}
\footnotemark[3]{\cite{Bodewits12b};}
\footnotemark[6]{Assumed;}
\footnotemark[5]{See text;}
\footnotemark[7]{comet X is a hypothetical comet similar to the observational 
condition of comet NEAT, but having
equal gas production rate of H$_2$O, CO$_2$, and CO; cal. = Calculated, obs. = Observation.}
\footnotemark[9]{Reference is for the observed G/R ratio value on the corresponding comet.}
\end{table*}

 \cite{Furusho06} observed the forbidden oxygen lines in comet \wild\ when it was 
at 2.4 au from the Sun. Using the infrared satellite AKARI, \cite{Ootsubo12} measured
the 
H$_2$O production rate in this comet as $\sim$1 $\times$ 10$^{27}$ s$^{-1}$ and 
 abundance of CO$_2$ was found to be  10\% relative to the water at heliocentric
distance of 2.22 au.
\cite{Ootsubo12} also determined the upper limit
for CO abundance in this comet as 20\% relative to water. In our model we have used 
these measured gas production
rates and relative abundances as input assuming that these values did not vary 
significantly in this comet
from 2.2 au to 2.4 au.

 Using the SPITZER space telescope, \cite{Woodward07}  measured the H$_2$O
production 
rate in comet \linear\
 as 2.43 $\times$ 10$^{29}$ s$^{-1}$ when the comet was at 1.76 au from the Sun
during 
pre-perihelion.
 \cite{Decock13}
observed atomic oxygen forbidden lines in this comet when it was at 2.6 au
heliocentric  distance.
Since the H$_2$O production rate was not measured at 2.6 au we scaled the 
\cite{Woodward07} measured
H$_2$O production rate to heliocentric distance of 2.6 au assuming that
it varies as the inverse square of heliocentric distance. However, we evaluate the
impact of the 
estimated H$_2$O production rate on the calculated G/R by decreasing its value by a 
factor of 2. Since CO$_2$ and CO are not observed in this comet we have assumed their
abundances to be 10\% and 25\% relative to H$_2$O, respectively. We show that CO
does not
play a significant role in determining green and red-doublet emission line
intensities,  whereas the CO$_2$ abundance is important in determining the \grat.

In comet \sid,  only the [OI] 6300 \AA\ emission line was  observed and
the intensity of [OI] 5577 \AA\ was estimated with a 3$\sigma$ upper limit
  when it  was at a heliocentric  distance of 2.96 au \citep{McKay12}.
 The AKARI satellite detected both H$_2$O and CO$_2$ 
in comet \sidq\ during its  
 pre-perihelion  period and  measured the CO$_2$ relative
abundance as 17\% relative to H$_2$O production rate when the comet was at a
heliocentric  distance of 3.3 au \citep{Ootsubo12}.
Assuming that the photodissociation of H$_2$O is the major source for the observed
[OI]  6300 \AA\ emission, 
 \cite{McKay12} inferred the H$_2$O production rate in comet \sidq\ as 1.8 $\times$
10$^{27}$ s$^{-1}$ which is smaller by a factor of 2 than the \cite{Ootsubo12} 
measurement. \cite{Ootsubo12} observation covers a larger (43$''$ $\times$ 43$''$)
projected  area on the coma
which can account for most of the H$_2$O produced from extended distributed
sources  like icy grains in the coma
 in the observed field of view 
 compared to that of the \cite{McKay12} observation (3.2$'' \times$ 1.62$''$).
Hence, we have used the \cite{Ootsubo12} measured gas production rates in the model.
We have
taken the H$_2$O production rate on comet \sidq\ as 4 $\times$ 10$^{27}$ s$^{-1}$
with 
17\% and 10\% relative abundances of  CO$_2$ and CO with respect to water, 
respectively, in our model.

By making radio observations on comet \chris, \cite{Bockelee10}  derived  H$_2$O and CO
 production rates as 4.2 $\times$ 10$^{28}$ and 3.9 $\times$ 10$^{28}$ s$^{-1}$, 
respectively. During this measurement the comet was at a heliocentric distance of 3.2
au. These values are higher by a factor of 2 compared to 
the infrared satellite observed values reported by \cite{Ootsubo12} which were 
derived when the comet was nearly at the same 
heliocentric distance. \cite{Ootsubo12} reported 42\% and 98\% of CO$_2$ and CO 
abundances relative to H$_2$O,
respectively, in this comet when it was at 3.13 au. During the green and red-doublet 
emission observation,
 comet \chrisw\ was at a heliocentric distance of 3.13 au \citep{McKay12}. For this
comet we have used the \cite{Ootsubo12} measured H$_2$O production rate as well as
CO$_2$ and CO relative abundances in our model.

The H$_2$O production rate in comet \garad\ beyond 2 au has been reported by various
workers \citep{Paganini12,Villanueva12,Bodewits12b,Combi13,Farnham12,Feaga12}.
Using the SWIFT satellite, \cite{Bodewits12b} observed the OH 3080 \AA\ emission line
in comet  \garadp\
 and derived the H$_2$O production rate when it  was between 2 au and 4 au
heliocentric distances. We have taken H$_2$O production at 3.25 au from the Sun as
2.3 $\times$  10$^{27}$ s$^{-1}$
by linearly interpolating the \cite{Bodewits12b} derived production rates
between  3 au and 3.5 au
heliocentric distances.
\cite{Decock13} used the observed \grat\ at 3.25 au and  estimated that
around 25\% CO$_2$ abundance relative to H$_2$O was present in this comet.
We assumed 25\% CO$_2$ relative abundance in the coma of comet \garadp\ in the model.
 Since CO is highly volatile and the comet is at a large heliocentric distance 
we assumed that the gas production rates for H$_2$O and CO are equal in this comet.

In comet \neat, the H$_2$O production rate is measured by \cite{Biver09} and 
\cite{Combi09}  at different heliocentric distances  using hydrogen Ly-$\alpha$ 
(1216 \AA) and radio (557 GHz) emissions, respectively. 
 \cite{Combi09} fitted the observed H$_2$O production rate as a function of heliocentric 
distance (r$_h$) as 3.5 $\times$ 10$^{29}$ $\times$  r$_h^{-1.7}$ s$^{-1}$. 
We used this expression to calculate the H$_2$O production rate
in this comet at 3.7 au where the green and red-doublet emissions were observed
\citep{Decock13}. Since the comet is at a large heliocentric distance we assumed that
the CO and H$_2$O abundances are equal. Based on the observed \grat\ on this comet,
\cite{Decock13} suggested that the CO$_2$ relative abundance in this comet could be 
between 60\% and 80\% with respect to H$_2$O. In our model we have assumed the CO$_2$
relative abundance at 3.7 au heliocentric distance to be 75\%.

To evaluate the individual contributions of major O-bearing species in producing
green  and red-doublet
emissions and their affect on the \grat\ we have made a case study for a hypothetical
comet X in which we assumed equal gas production of H$_2$O, CO$_2$, and CO in the comet.
This is similar to the observation of \cite{Ootsubo12} on  comet \chrisw\ in which it
is found that CO$_2$ and H$_2$O gas production rates  are equal
 ($\sim$8 $\times$ 10$^{27}$ s$^{-1}$); however, the CO production rate is around  3
times higher  when the comet was at 3.7 au from the Sun.

The solar flux, which is required to calculate photorates of different species,  is
taken from the SOLAR2000 (S2K) v.2.36 model
 of \cite{Tobiska04} at 1 au and scaled  according to the observed
heliocentric distance of different comets.
The electron temperature that determines the
dissociative recombination rates of ions is taken as constant 300 K in the cometary
coma. The effect of this constant
temperature profile on the model calculation is discussed later.
The yield of \ols\ at solar H Ly-$\alpha$ in the photodissociation of H$_2$O is taken
 as 0.5\%.
The impact of this assumption was discussed in our previous work \citep{Bhardwaj12}.
The photodissociative excitation cross section for CO$_2$ producing \old\ is taken from
\cite{Jain12b}. The photorate for the production of \ols\ from the photodissociation of 
CO has been taken from  the theoretically estimated value of \cite{Festou81} and scaled  
to the observed heliocentric distance. 

\section{Results}
Since these comets have different water production rates (varying from 10$^{27}$ to 
10$^{29}$ s$^{-1}$), as well as different CO$_2$ and CO relative abundances with
respect to H$_2$O,  we present calculations in comet \chrisw\ which is followed by a
discussion on the calculated results for other comets.

\subsection{Production processes of \ols\ and \old}
\begin{figure}
\begin{center}
\includegraphics[width=22pc]{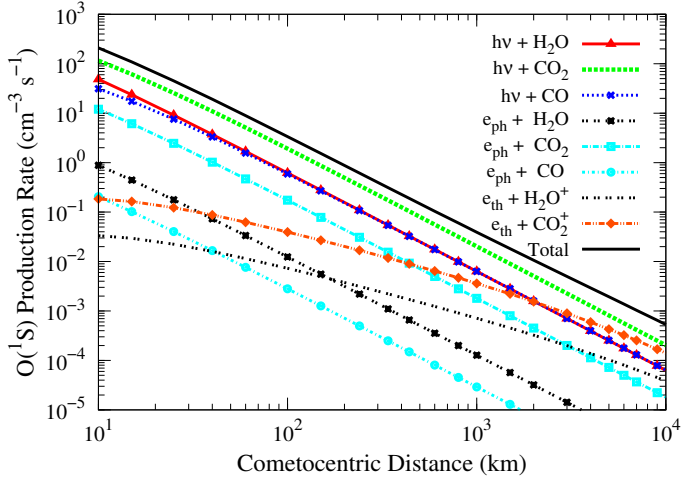}
\caption{Calculated volumetric \ols\ production rate profiles for major production 
mechanisms in comet \chris\ with H$_2$O production rate of 2 $\times$ 10$^{28}$
s$^{-1}$ and  42\% CO$_2$ and  98\% CO abundances relative to H$_2$O in the cometary
 coma when the comet was at 3.13 au heliocentric distance. h$\nu$: solar photon;
e$_{ph}$: photoelectron; and  e$_{th}$: thermal electron.}
\label{fig:o1s-prod}
\end{center}
\end{figure}

The calculated production rates for the \ols\ from different  processes 
 in comet \chrisw\ are presented in 
Fig.~\ref{fig:o1s-prod}. The major production source of oxygen atoms in the $^1$S 
metastable state is photodissociation of CO$_2$ followed by photodissociation of CO
and H$_2$O. The contribution from the photoelectron impact excitation reactions is smaller 
compared to photodissociative excitation processes. Above 10$^3$ km, the
dissociative 
recombination of CO$_2^+$ also contributes significantly.  The solar flux in
wavelength bin 955--1165 \AA\ is the main source that dissociates CO$_2$ and
produces atomic oxygen in the $^1$S state. Since the yield for photodissociation of
CO$_2$ in this wavelength bin is almost unity, the absorption of solar photons of
this  wavelength bin by CO$_2$ leads to the formation of O($^1$S) and CO
\citep[][]{Raghuram13,Bhardwaj12}.

\begin{figure}
\begin{center}
\noindent\includegraphics[width=22pc,angle=0]{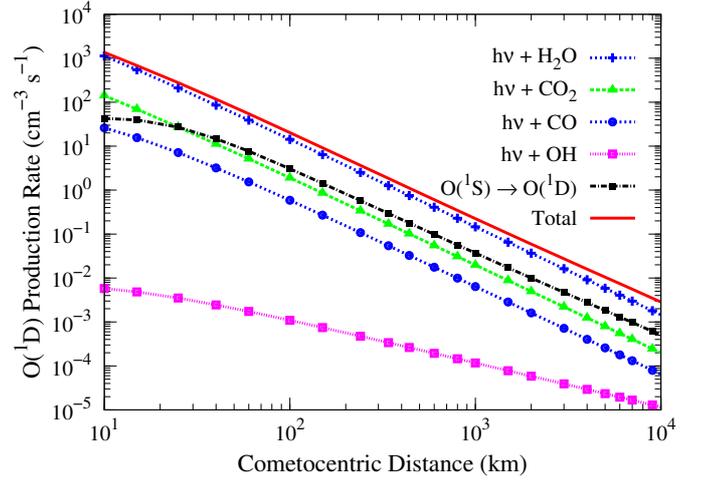}
\caption{Calculated volumetric \old\ production rate profiles for major production 
mechanisms
in comet \chris\ with H$_2$O production rate of 2 $\times$ 10$^{28}$ s$^{-1}$ and
 42\% CO$_2$ and  98\% CO relative abundances with respect to H$_2$O 
 in the cometary coma when the comet was at 3.13 au from the Sun.
  h$\nu$: solar photon.}
\label{fig:1d-prod}
\end{center}
\end{figure}

The calculated \old\ production rate profiles for different mechanisms are shown
in  Fig.~\ref{fig:1d-prod}. The major production of \old\ is via
photodissociation of H$_2$O, but close to the  nucleus ($<$30 km)
photodissociation of CO$_2$ is also  a significant \old\ production process, and
above 30 km the radiative decay of \ols\ becomes a more important source 
of \old\ than the former. The photodissociation of CO and OH are minor production
sources  of \old.
 Most of the \old\ production ($>$95\%) is due to  photodissociation
of H$_2$O  by solar H Ly-$\alpha$ photon flux.

\subsection{Loss processes of \ols\ and \old}
\begin{figure}
\begin{center}
\noindent\includegraphics[width=22pc,angle=0]{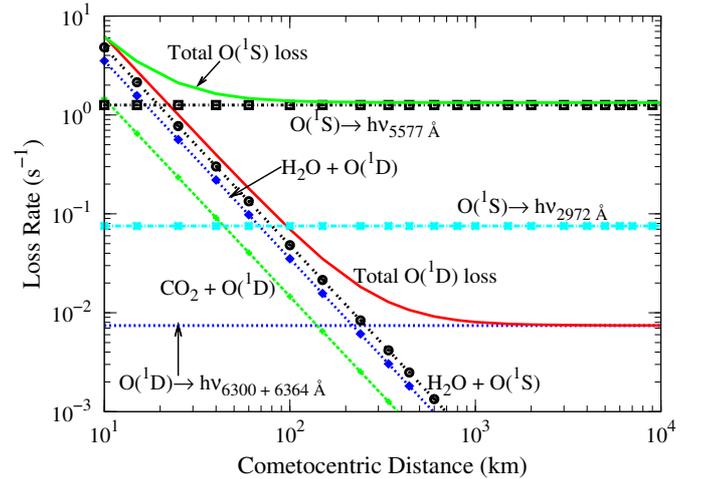}
\caption{Calculated radial loss rate profiles for major loss mechanisms of the \old\ 
and \ols\ 
in comet \chris\ with H$_2$O production rate of 2 $\times$ 10$^{28}$ s$^{-1}$ and
 42\% CO$_2$ and 98\% CO abundances relative to H$_2$O 
 in the cometary coma when the comet was at 3.13 au
from the Sun.}
\label{fig:1ds-los}
\end{center}
\end{figure}

 The calculated \ols\ and \old\ destruction rate profiles in comet \chrisw\ are
presented in Fig.~\ref{fig:1ds-los}. 
Since this comet has a low neutral gas production rate, the collisional quenching is a
dominant \ols\ destructive mechanism only close to the nucleus ($<$30 km).
The radiative decay  which produces photons at 5577 \AA\ and 
2972 \AA\ is the major loss process for \ols\ throughout the coma.
The calculated loss rate profiles of \old\ by various processes are also presented in
the same figure. Below 300 km, quenching by H$_2$O and CO$_2$
are the dominant loss mechanisms of the \old. Above 300 km, the 
radiative decay,  which results in the emission of 
photons at 6300 \AA\ and 6364 \AA, is a major loss process for \old.  Quenching 
by CO
is a minor loss process for \old\ which is not shown in the figure.

\begin{figure}
\begin{center}
\noindent\includegraphics[width=22pc,angle=0]{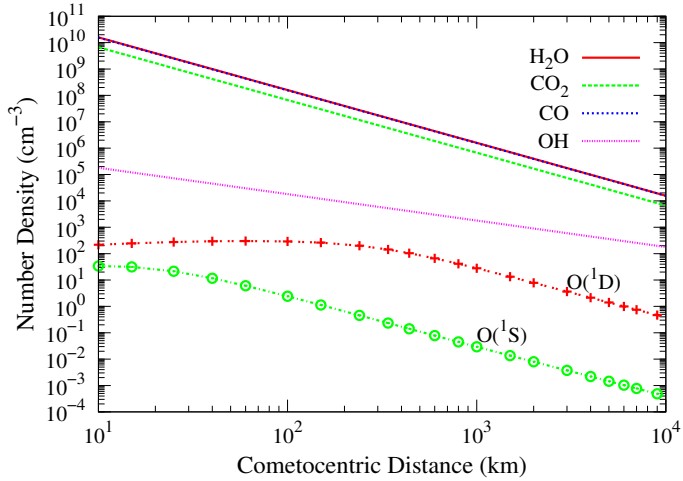}
\caption{Calculated number density profiles of \ols, \old, 
 and OH, along with those of H$_2$O, CO, and CO$_2$ in 
comet \chris\ with H$_2$O production rate of 2 $\times$ 10$^{28}$ s$^{-1}$
and 42\% CO$_2$ and 98\% CO relative abundances with respect to H$_2$O 
 in the cometary coma when the comet was at 3.13 au from the Sun.}
\label{fig:nden}
\end{center}
\end{figure}

The calculated number density profiles of \ols\ and \old\ in comet \chrisw\ along
with  parent species H$_2$O, CO$_2$, and CO are presented in Fig.~\ref{fig:nden}.
Close to the cometary nucleus the flatness in the calculated \ols\ and \old\ number
density profiles is due to collisional quenching by cometary species (mainly H$_2$O)
and depends on the neutral gas production rate of the comet.

 \subsection{[OI] green to red-doublet emission intensity ratio and line widths}
\begin{figure}
\begin{center}
\noindent\includegraphics[width=22pc,angle=0]{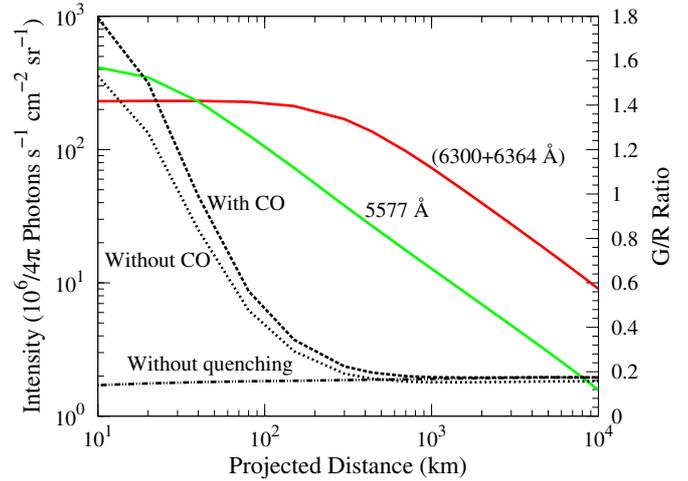}
\caption{Calculated [OI] red-doublet (6300+6364 \AA) and 5577 \AA\ line brightness
profiles (with solid curves) along the cometocentric projected distances on comet 
\chris\ with H$_2$O production rate of 2 $\times$ 10$^{28}$ s$^{-1}$ and 
 42\% CO$_2$ and 98\% CO relative abundances with respect to H$_2$O in the cometary 
coma when the comet was at 3.13 au from the Sun. The calculated G/R ratio profiles
 considering CO, without considering CO, and without quenching are plotted
 with dashed, dotted, and dash-dotted curves, respectively, on the
right y-axis.} 
\label{fig:rg-inten}
\end{center}
\end{figure}

The calculated number density profiles shown in Fig.~\ref{fig:nden} are multiplied
with Einstein emission transition probabilities \citep[see Table 1 in][]{Raghuram13}
to obtain emission rates. By integrating these emission rates along the line of sight
we calculated the emission intensities of green and red-doublet lines as a function
of projected distance.
The calculated surface brightness profiles for [OI] 5577 \AA\ and red-doublet  (6300
+ 6364 \AA) emissions
 are shown in Fig.~\ref{fig:rg-inten} with solid curves. 
It can be seen in this figure that close to the nucleus  
(below 40 km projected distance) the green line emission is more intense than the 
red-doublet emission,
which is mainly due to the \ols\ emission rate (1.26 s$^{-1}$) being higher by about 
two orders of magnitude
compared to that of \old\ (8.59 $\times$ 10$^{-3}$ s$^{-1}$).
The calculated \grat\ in comet \chrisw, which is shown with  a dashed curve (''with
CO'')  in Fig.~~\ref{fig:rg-inten},
varies between 1.8 and $\sim$0.2.  In the same figure  the calculated \grat\ profiles
for different cases are also presented.
Since there is an uncertainty in the photo-rate of CO in producing \ols, which is
discussed later,  we also did calculations for the \grat\ neglecting this source mechanism
which is shown in Fig.~\ref{fig:rg-inten} with dotted curve (''without CO'').
 In this case the calculated \grat\ varies between 1.6 and 0.18.  Since comet
\chrisw\ has a very low gas production rate the collisional quenching may be less
important.
To assess the effect of collisional quenching  on the green and red-doublet
emissions, we calculated the \grat\ without considering collisional destruction
mechanisms  of \ols\ and \old. In this case the calculated \grat\ is a constant value
of 0.18 throughout the coma which is represented with the dash-dotted line in
Fig.~\ref{fig:rg-inten}.

\begin{figure}
\begin{center}
\noindent\includegraphics[width=22pc,angle=0]{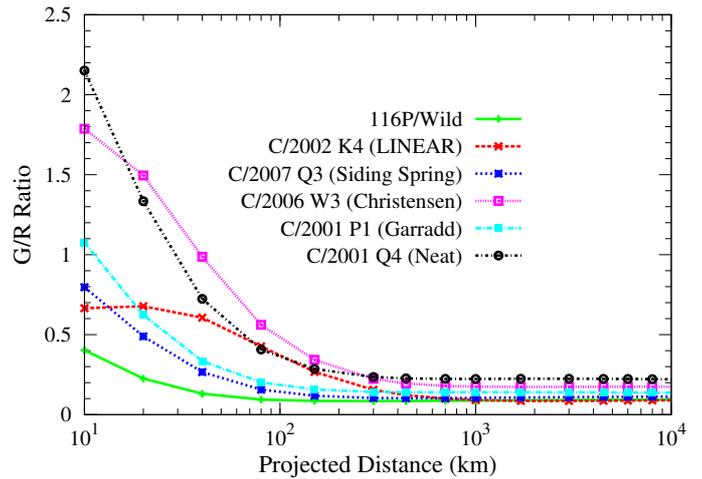}
\caption{The calculated radial profiles of the \grat\ in different comets. The input 
parameters used to calculate the \grat\ are tabulated in Table~\ref{tab-comet}. It
can be seen that in comets \chrisw\ and \neatq, due to substantial collisional
quenching of \ols\ and \old\ with other cometary species, the calculated
 \grat\ is more than 1 closer to the nucleus, whereas above 400 km projected
distances
 it is a constant.}
\label{fig:grat-prj}
\end{center}
\end{figure}

Similarly, all these calculations have been carried out on other comets.
Considering both collisional quenching and 
photodissociation of CO, the calculated \grat\ profiles as a function of projected 
distance in six comets are presented in Fig.~\ref{fig:grat-prj}. This figure shows
that close to the nucleus in comets \chrisw\  and \neatq, the
calculated \grat\  value is more than one which is due to higher CO$_2$ relative
abundances and strong collisional quenching of \old\ by cometary species, whereas in
other comets this value is always less than one throughout the coma. In comets
\chrisw\ and \neatq, the CO$_2$ abundances are very large (see Table~\ref{tab-comet})
and no significant collisional quenching of \ols. Thus, the  green line
intensity throughout the coma is determined by CO$_2$ and subsequently the \grat\
governed  by the quenching of \old\ depending on the H$_2$O production rate. In other
comets the \grat\ is small because of lower CO$_2$ abundances compared to former
comets.

We calculated the average \grat\ over
the observed projected area on each comet. The  projected area on a comet
changes with the dimension of slit used for observation and the geocentric distance
of  comet.
The calculated averaged G/R ratios
on different comets are tabulated in Table~\ref{tab-comet} along with the values
derived  from observations.
 Our calculated \grat\ values are comparable (within a factor of two) with the 
observations on
different comets observed at different
heliocentric and geocentric distances.
 
The percentage contributions for various production processes involved in the formation
of \ols\ and \old\ in these comets at three different projected distances are
given in Table~\ref{tabprj}. These calculations suggest that in all
these comets, the photodissociation of H$_2$O and CO$_2$ together 
 produce 50--80\% of \ols, whereas, irrespective of CO$_2$ and CO relative abundances,
the major ($\sim$50 to 80\%)  source for the formation of \old\ is photodissociation 
of H$_2$O  
followed by radiative decay of \ols\ (10--15\%). At larger  projected distances 
($>$10$^3$ km), 
dissociative recombination processes of H$_2$O$^+$ and CO$_2^+$ ions are also important
 production sources of 
\ols\ (30--40\%) and \old\ ($\sim$20\%). 

\begin{table*}[tbh]
\renewcommand{\thefootnote}{\fnsymbol{footnote}}
\begin{center}
\caption{Calculated percentage contributions for major production processes of 
O($^1$S) and O($^1$D) in different comets.} 
\scalebox{0.8}[0.9]{
\label{tabprj}

\begin{tabular}{@{}lcccccccccccccccccccc@{}}

\toprule
\multicolumn{1}{c}{} &  
\multicolumn{18}{c}{Production processes of O($^1$S) and O($^1$D) at three
 cometocentric projected distances (km) (\%)}    \\ % [2pt]
%  \cline{2-19}
% \cline{1-2}  \cline{2-8}
Comet&\multicolumn{3}{c}{ h$\nu$ + H$_2$O}
&\multicolumn{3}{c}{ h$\nu$ + CO$_2$}  
&\multicolumn{3}{c}{ h$\nu$ + CO}
&\multicolumn{3}{c}{ O($^1$S) $\rightarrow$  O($^1$D)} 
&\multicolumn{3}{c}{ e$_{th}^-$ + H$_2$O$^+$  } 
&\multicolumn{3}{c}{ e$_{th}^-$ + CO$_2^+$} \\ [2pt]
% \cline{2-19}
\cmidrule{2-4} \cmidrule{5-7} \cmidrule{8-10} \cmidrule{11-13} \cmidrule{14-16} 
\cmidrule{17-19}
 &10$^2$&10$^3$&10$^4$&10$^2$&10$^3$&10$^4$&10$^2$&10$^3$&10$^4$&10$^2$&10$^3$&10$^4$
 &10$^2$&10$^3$&10$^4$&10$^2$&10$^3$&10$^4$\\ [5pt]
\midrule
\wild   & 43     & 33   & 25   & 33   & 25    &  20   & 9   &
7   & 5    & & &   & 5     & 15   & 22  & 5   & 15  & 17   \\  
   &   (83)\footnotemark[1]  &   (71) &   (59) &   (3)  &   (2)  &    (2) &   (0.5) &
  (0.5) &   (0.5)  &(8)&(9)&(9)  &   (4)  &  (14) &  (22) &   (0.1) &   (0.7)&    (2)
 \\ [8pt]
\linear & 46     & 42     &  32   & 35   & 32  & 25 & 11  & 11  & 9   
& & &   & 1   & 4  & 13   & 1   & 5  & 14   \\
  &    (88)  &  (83) &  (71) & (3) &  (3) &  (2)&  (1) &  (1)&  (1)
&(7)&(8)& (9)  & (0.5) &  (3) &  (13) &  (0.5) &   (0.5) &   (1) \\   [8pt]
\sid    &  35    & 29   & 21    & 46   & 37  & 28  & 4   & 3 & 2   
&  &  &   & 2   & 18  & 17   & 4   & 15   & 23   \\
    &  (82)  &  (73) &   (60) &  (4) & (4) &  (3)&  (0.5) &   (0.5)&   (0.5)
&(9)&(10)&(11) &  (2) &   (9) &  (20) &   (0.5) & (2) &  (3) \\  [8pt]
\chris & 18   & 15  & 11   & 55  & 47  & 37 & 17   & 15  & 11  
&   &   &   & 0.5   & 3   & 8   & 3   & 15   & 28  \\
  &   (69) &  (61) & (48) &  (9) &   (8)&   (7)& (3) &  (2)&   (2)
& (15) & (16) &   (17) &  (1) &  (6) &   (15) &   (0.5) &   (3) &   (5) \\  [8pt]
\garad & 22    & 17   & 14  & 41  & 32  & 28  & 22  & 17 & 15 
&   &  &   & 3   & 9   & 13  & 8  & 20  &  23  \\
 &   (71)  &  (58) &   (47) &   (6) &   (5) & (4) &   (3) &   (3)&    (2)
& (13) &(14) & (13) &   (4) &  (14) &  (19) &  (1) &   (3) &    (3) \\ [8pt]
\neat & 11   & 9   & 7   & 63   & 50   & 40  & 11   & 9  & 7  
&  &  &   & 1  & 4  & 6  & 7 & 23  &  34  \\
  &   (57) &   (47) &  (37) &   (13) &   (11) &   (9)&  (2) &   (2)&    (2)
& (21) & (22) & (22) &   (2) &   (8) &   (14) &   (1) &   (5) &   (8) \\  [10pt]

comet X\footnotemark[2] & 8   &  6  & 5   & 63   & 47 & 41  &  8  & 6  & 5
 &  &  &   & 1  & 3  & 5  & 12 & 30  & 37   \\
  &   (48) &   (38) &  (31) &  (15) & (12) & (10)& (2) &   (1)&    (1)
& (23) & (25) & (24) &   (3) &   (10) &   (13) &   (3) &   (8) &   (10) \\ % [10pt]
\bottomrule
\end{tabular}}
\end{center}
\footnotemark[1]{The values in parentheses are for \old;}
\footnotemark[2]{comet X is a hypothetical comet similar to the observational 
condition of comet NEAT and having
equal gas production rates of H$_2$O, CO$_2$, and CO; h$\nu$ = photon;  e$_{th}^-$ =
thermal electron.}
 \end{table*}
%  \end{table}

The calculated percentage contributions of different production processes in the 
total intensity of [OI] emissions over the observed coma on these comets 
are tabulated in Table~\ref{tab-slit}.
These calculations suggest that photodissociation of CO$_2$ and H$_2$O together 
contribute 50--70\% to the green line emission and the remaining contribution is
through dissociative recombination  of H$_2$O$^+$ and CO$_2^+$ ions. In the case of
red-doublet emission, photodissociation of H$_2$O and radiative decay of \ols\
together produce 70--90\% and contributions from other sources are very small.

In the case of hypothetical comet X which has equal H$_2$O, CO$_2$, and CO, 
gas production rates, 
$\sim$80\% of green line emission intensity is governed by CO$_2$  (via
photodissociation of CO$_2$ and
 dissociative recombination of CO$_2^+$), whereas
photodissociation of H$_2$O and CO together contribute around 10\%. Dissociative
recombination of CO$_2^+$ is the second important source and contributes around
30\% to the total green line emission. In this case around 35\% of red-doublet
emission is produced
via H$_2$O photodissociation. The production of \old\ via CO$_2$
photodissociation is around 10\% of the total while it is $\sim$25\% via radiative 
decay of \ols, which is also
essentially produced from CO$_2$. In this case both CO$_2$ and H$_2$O play 
equally important roles in producing red-doublet emission. 

 We also calculated the mean excess energy released in these 
photodissociative excitation reactions. The maximum excess energy in photodissociation
of H$_2$O producing \ols\ by solar Ly-$\alpha$ photons is 1.27 eV, whereas the mean 
excess energy
in the photodissociation of CO$_2$ forming \ols\ is 2.55 eV. Mean excess energies
in photodissociative excitation of H$_2$O, CO$_2$, and CO producing \old\
 are 2.12, 4.46, and 2.54 eV, respectively.
We assumed that most of these excess energies will result in kinetic motion of
daughter products. Thus, the excess velocities of \ols\ in photodissociative 
excitation of H$_2$O and CO$_2$ are 1.3 km s$^{-1}$ and 4.4 km s$^{-1}$,
respectively. Similarly, the calculated  excess velocities of \old\ in
photodissociation of H$_2$O, CO$_2$, and CO are 1.6, 5.8, and 3.6 km
s$^{-1}$,  respectively.

Considering only photoreactions and using the calculated
contributions of each process over the cometary coma (see Table~\ref{tab-slit})
 we calculated the mean excess energies of \ols\ and \old.
Our calculated mean velocities of \ols\ and \old\ atoms on these comets
 are tabulated in Table~\ref{tab-slit} along with 
the derived velocities based on the observed line widths.
In comets having large CO$_2$ relative abundances the width of the green line, which
is a
 function of 
mean \ols\ velocity,  is mainly determined by photodissociation of CO$_2$.
Since the mean excess energy released in photodissociation of CO$_2$ is higher, the
width of the green line would be greater compared to the red-doublet emission line
width
(which is mainly determined by photodissociation of H$_2$O).
Our calculated green line widths in different comets, which are presented in 
Table~\ref{tab-slit},
are higher than the calculated red-doublet emission line widths, which is 
consistent with the observations.

\renewcommand{\thefootnote}{\fnsymbol{footnote}}
\begin{table*}[tbh]
\begin{center}
 \caption{Calculated percentage contributions for major production processes
 of green and red-doublet emissions in the slit projected field of view on different 
comets,
and the comparison of the calculated and observed line widths.}
\scalebox{0.8}[0.95]{
\label{tab-slit}

\begin{tabular}{lcccccccccccccccccccccccc} 
\toprule
 \multicolumn{1}{l}{Comet} 
&\multicolumn{1}{c}{h$\nu$ + H$_2$O} 
&\multicolumn{1}{c}{h$\nu$ + CO$_2$}
&\multicolumn{1}{c}{h$\nu$ +  CO}
&\multicolumn{1}{c}{O($^1$S) $\rightarrow$ O($^1$D)}
&\multicolumn{1}{c}{e$_{th}^-$ + H$_2$O$^+$}
&\multicolumn{1}{c}{e$_{th}^-$ + CO$_2^+$} & \multicolumn{2}{c}{5577 Line width} & 
\multicolumn{2}{c}{6300 Line width}\\[2pt]
\midrule
      &  & & && & & Cal\footnotemark[3] & Obs\footnotemark[2] & Cal\footnotemark[3] &
  Obs\footnotemark[2]  \\
\cmidrule{8-11}
\wild      & 34 (72)\footnotemark[1] & 26 (2)& 7 (0.5)&(9)& 14 (13)& 14 (1)& 1.70  & --          & 1.32  & --         \\[2pt]
\linear    & 40 (81) & 31 (3)& 10 (1) &(8)& 6 (5)& 7 (0.5) & 2.05  & 2.38--2.76  & 1.87  & 1.81--2.12 \\[2pt]
\sid       & 28 (72) & 37 (4)& 3 (0.5)&(10)& 9 (10)& 15 (1)& 2.04  & --          & 1.44  & --         \\[2pt]
\chris     & 14 (61) & 47 (8)& 14 (3) &(16)& 3 (6)& 15 (3) & 2.48  & --          & 1.58  & --         \\[2pt]
\garad     & 16 (55) & 31 (4)& 17 (3) &(13)& 10 (15)& 21 (3) & 1.85  & 2.16--2.54  & 1.25  & 1.25--1.67 \\[2pt]
\neat      & 8  (44) & 47(11)& 8 (2)  &(22)& 4 (9)& 26 (6)  & 2.30  & 2.31--2.55  & 1.65  & 2.39--2.75 \\[2pt]
comet X\footnotemark[4] & 6  (36) & 46(11)& 6 (1)  &(25)& 4 (10)& 31 (8)  & 2.35 & --  & 1.65  & -- \\
\bottomrule
\end{tabular} 
}
\end{center}
\footnotemark[1]{The values in parentheses are calculated percentage  contributions
for 
red-doublet emission;}
\footnotemark[2]{Obs: observed line widths are taken from \cite{Decock13};}
\footnotemark[3]{Cal: model calculated line widths;}
\footnotemark[4]{comet X is a hypothetical comet similar to the observational
condition  of comet NEAT and having
equal gas production rates of H$_2$O, CO$_2$, and CO; h$\nu$ = photon; e$_{th}^-$ = 
thermal electron.}
\end{table*}

\begin{center}
\begin{figure}
\noindent\includegraphics[width=22pc,angle=0]{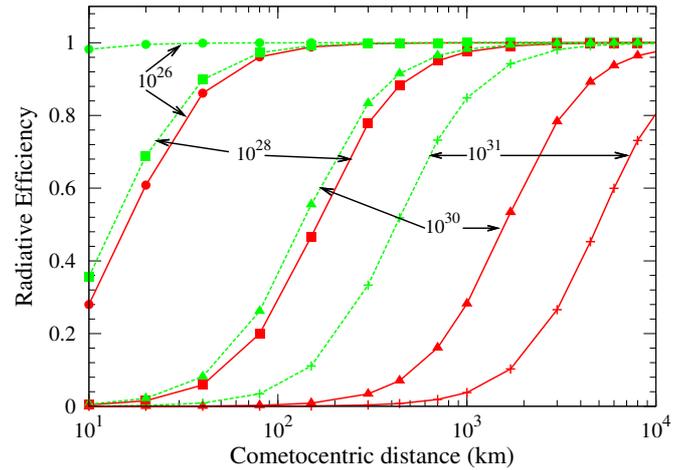}
\caption{The calculated radiative efficiency profiles of \ols\ (green lines) and
\old\  (red lines) for different production rates with 5\% CO$_2$ and 10\% CO
relative abundances with  respect to H$_2$O.  The circles, squares, triangles, and
cross symbols represent the calculated radiative efficiencies for the water
production  rates of 10$^{26}$,  10$^{28}$,  10$^{30}$, and  10$^{31}$,
respectively.}
\label{fig:rad-eff}
\end{figure}
\end{center}

Depending on the composition and activity of the  nucleus, comets have different
gas production rates at different heliocentric
distances. In order to appraise the collisional quenching of \ols\ and \old\  with
increase in H$_2$O production rate,  we calculated the radiative efficiencies of
\ols\ and \old\ for different water  production rates. The calculated radiative
efficiency profiles are shown in Fig.~\ref{fig:rad-eff}.  This calculation
shows that for a given water production rate  the \old\ is always much more 
quenched  than that of \ols. This is mainly because the lifetime of \old\ ($\sim$120
s) is larger by two orders of  magnitude than that  of \ols\ ($\sim$0.8 s).

\begin{center}
\begin{figure}
\noindent\includegraphics[width=22pc,angle=0]{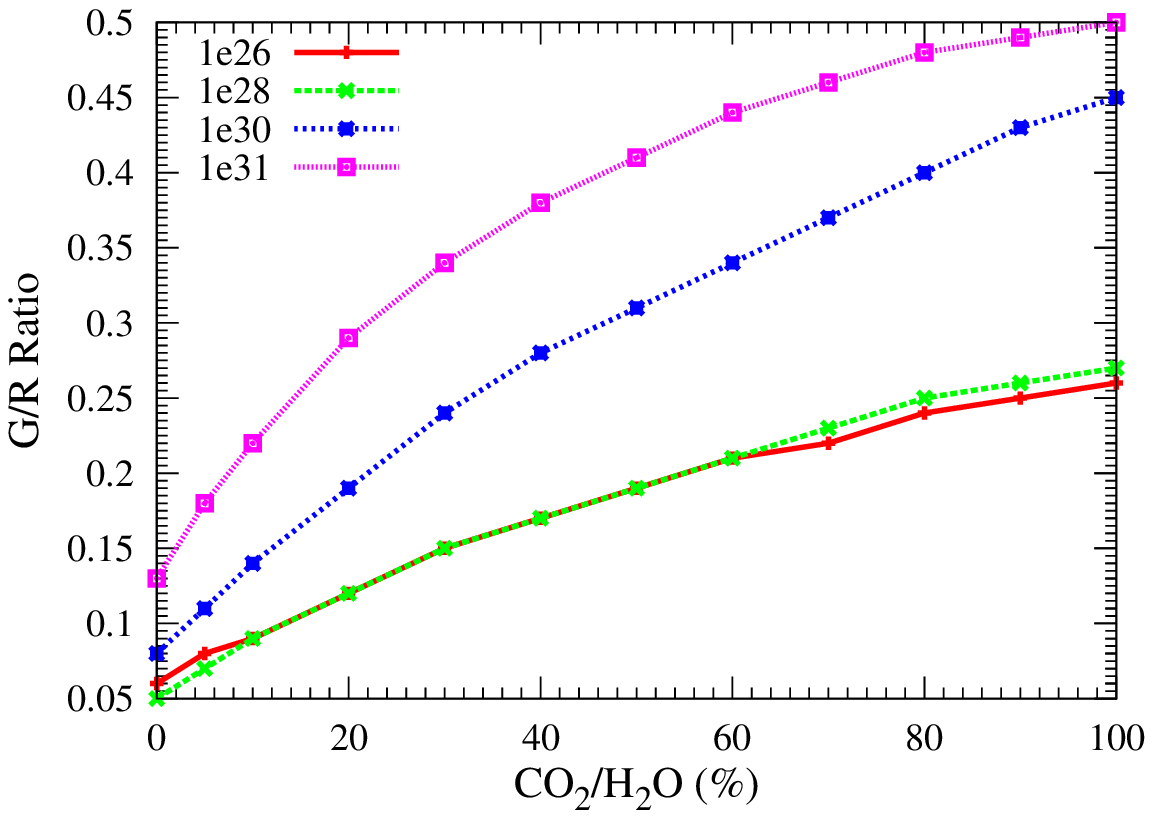}
\caption{The calculated \grat\ as a function of CO$_2$ abundance
 for different water production rates, with 10\% CO abundance relative to H$_2$O in 
the coma.
These calculations are done at heliocentric  and  geocentric  distances
 of 1 au using a square slit of side 5$''$.}
\label{fig:gr-co2}
\end{figure}
\end{center}

Since CO$_2$ is a potentially  important source of \ols\ we have also carried out
model  calculations of the \grat\ on different water  production rates by varying its
relative abundance  from 0\% to 100\% with respect to H$_2$O. The calculations
presented in Fig.~\ref{fig:gr-co2}  suggests that  by increasing CO$_2$
relative abundance in a comet the \grat\ increases almost monotonically.

\section{Discussion}
For comets close to 1 au  from the Sun the dominant species in the cometary coma is
H$_2$O.
Because of  lower ice-sublimation temperatures of CO and CO$_2$, at large
heliocentric
distances the cometary coma is dominantly composed of CO$_2$ and CO
\citep{Meech04,Crovisier99,Biver97,Biver99,Bockelee04,Bockelee10}.
Owing to strong absorption of cometary H$_2$O infrared emission lines by terrestrial 
water molecules,
it is  difficult to detect H$_2$O in the coma for ground-based observations, but
the spatial profiles of water can be easily derived in comets from ground-based 
observatories by observing
 infrared H$_2$O non-resonance fluorescence emissions
 \citep{Mumma95,Mumma96,Russo00}.
 Since H$_2$O does not have any transitions in the visible region                    
the emissions of its daughter products have been used as tracers to understand the 
spatial distribution of water in the cometary coma. 
The observed atomic oxygen visible emissions  
(viz. O[I] 5577, 6300, and 6364 \AA) have been used to quantify the H$_2$O
production  rate in several
comets around 1 au \citep{Delsemme76,Delsemme79,Fink84,Schultz92,Morgenthaler01}.
 Since CO$_2$ and CO can also produce these metastable oxygen atoms,
based on the theoretical work of \cite{Festou81}, the \grat\ of 0.1
 has been used as the benchmark to confirm H$_2$O as the parent species for these
oxygen  emission lines.
The available theoretical and experimental cross sections for the
production of \ols\ and \old\ from different O-bearing species have been reviewed in our
previous work  \citep{Bhardwaj12}.
Our coupled chemistry-emission model,
 which has been applied to comets \hya\ and \hbop, suggested that the observed
\grat\ not only depends on the relative abundances of CO$_2$ and CO, but also on the 
projected area
observed on the comet \citep{Bhardwaj12}.

Since CO$_2$ does not emit ultraviolet
or visible photons we cannot detect this molecule directly in the cometary
ultraviolet  or visible spectra.
 Moreover, CO$_2$ is a symmetric molecule with no permanent dipole moment
and so it is difficult to observe this molecule even in radio range from the
ground  \citep{Ootsubo12}.
Thus, this molecule is probed using indirect methods using the emissions of  its
dissociative
products, like the CO Cameron band (a$^3\Pi$--X$^1\Sigma^+$) in ultraviolet 
\citep{Weaver94,Feldman97}
and visible atomic oxygen green and red-doublet emissions 
\citep{Furusho06,McKay12,Decock13}. Our
earlier works \citep{Bhardwaj11,Raghuram12} have shown that CO Cameron band emission 
is not suitable for measuring CO$_2$ abundances in comets since this
emission is mainly
governed by photoelectron impact excitation of CO rather than the photodissociation
of  CO$_2$.

Assuming that the green line emission is governed by photodissociation of   CO$_2$
while the 
red-doublet emission is controlled by photodissociation of H$_2$O, the observed
\grat\  has been used to
quantify CO$_2$ relative abundance in the comets \citep{McKay12,Decock13}. 
At larger heliocentric distances CO$_2$ and CO are the dominant O-bearing species in 
the coma which can
produce green and red-doublet emissions. In several comets  the observed  \grat\
 at large ($>$2 au) heliocentric distances is more than 0.1 
\citep{Decock13,McKay12,Furusho06}.

\subsection{Impact of CO on the \grat}
\label{sec:roco}
At larger heliocentric distances, although CO abundance is substantial in the
cometary 
comae,
the  photodissociation of CO is not a  potential source of \ols\ and \old\ atoms.
This is mainly due to the proximity  in the threshold energies 
of photodissociative excitation and photoionization of CO molecules.
The threshold energies for dissociation of CO into \old\ and \ols\ states are 14.35 
and 16.58 eV,
respectively, whereas it is 14 eV for ionization. 
 Moreover, the branching ratio of ionization for
 the photons having energy more than 14 eV is $\sim$0.98 \citep{Huebner92}. Since the
ionization energy is smaller than energy required in the formation of \ols\ and \old,
 most of the
photons ($>$90\%) having energy $>$14 eV ionize the CO molecule rather than causing 
photodissociative
excitation.
Based on  \cite{Huebner79} compiled cross sections, \cite{Festou81} estimated that 
the photodissociation of CO produces \ols\ and \old\ with nearly equal rates.
To evaluate the role of CO we also did calculations
 in comet \chrisw\  by discarding photodissociation of CO as a
source mechanism of both \ols\ and \old\ (see Fig.~\ref{fig:rg-inten}).
Though CO production rate is equal to that of H$_2$O in this comet 
(see~Table~\ref{tab-comet}), by removing
CO contribution the calculated \grat\ decreased by a maximum of about 10\%.
Similarly, our calculated percentage contribution over the observed 
coma on different comets,
which is presented in Table~\ref{tab-slit}, also suggests that the role of CO is 
very small ($<$20\%)
 in producing 
\ols\ and \old\ atoms and subsequently in  determining the red-doublet emission intensity. 
Even without considering photodissociation of CO in the model the calculated \grat\ values
 are in agreement with the observations.
Based on these calculations we can suggest that the photodissociation of
CO is an insignificant source of \ols\ and \old.
Hence, the photodissociation of CO has almost no impact on the \grat.

\subsection{Impact of CO$_2$ on the \grat}
The relative abundance of CO$_2$ with respect to H$_2$O, is very important in 
determining the \grat. This can be understood from the calculated 
 \grat\ profiles on comets \wildp, \sidq, and \neatq, which have
nearly same H$_2$O production rates (1--4 $\times$ 10$^{28}$ s$^{-1}$), but different
relative abundances of CO$_2$ and CO, and are observed at different heliocentric
distances (see Fig.~\ref{fig:grat-prj}). As discussed in Section~\ref{sec:roco},  the
CO abundance does not have
any appreciable impact on the \grat. Hence,  the change in the calculated \grat\
 on these comets can be ascribed mainly to the difference in CO$_2$
relative abundances.  The calculated \grat\ profiles on these comets are shown in 
Fig.~\ref{fig:grat-prj}, which  shows that by  increasing CO$_2$ the \grat\
increases. In comet \neatq,  due to higher (75\%)
 CO$_2$ relative  abundance,  the calculated  \grat\ value is more than one close to
the cometary nucleus. Similar behaviour is  seen for comet \chrisw\ which is due to
larger ($\sim$40\%)  CO$_2$ relative abundance with respect to H$_2$O, 
and also due to significant collisional quenching of \old\ (see
~Fig.~\ref{fig:rg-inten}). We found that by doubling  CO$_2$ relative abundance the
\grat\ changes by $\sim$25\%, whereas collisional quenching alone can vary its value
by  an order of magnitude. The model calculated \grat\ values in comets \wildp,
\sidq, and \garadp\ are smaller by a factor of around 1.5 compared to the
observations.

The detection of CO$_2$ molecules in the coma has been carried out using several
infrared satellites by observing its fundamental vibrational band emission ($\nu_3$) 
at 4.26 $\mu$m
 \citep{Crovisier96,Crovisier97,Crovisier99,Colangeli99,Reach10,Ootsubo12}.
The quantification of CO$_2$ abundance based on the observed infrared  emission 
intensity is
 subjected to opacity of the cometary coma. Since the CO$_2$ fluorescence efficiency
factor ($g-$factor)  is larger compared to that of H$_2$O and CO \citep[see Table 2
of ][]{Ootsubo12}, these
 emission lines are optically thick in the inner coma, which can result in
 underestimation of CO$_2$ abundance if proper treatment of radiative transfer
 is not accounted for in the analysis.
 The optical depth effects in the inner coma 
 may cause the surface brightness profile of these emissions to be much flatter and 
resemble the presence of extended sources in the coma. In comet Hale-Bopp,
 \cite{Bockelee10} have shown that the observed broad extent of infrared  CO
brightness
is due to optical depth effects of the emitted radiation and not because of
extended sources.
Since these comets are observed at larger heliocentric distances and have low gas
production rates, the collision dominated coma size is only a few
hundred kilometers. Thus, the opacity effects of these IR emissions can be 
significant close to the nucleus and can influence the derivation of CO$_2$
production
rate based on the observed flux over the field of view. The discrepancies between the
\cite{Ootsubo12}
 derived production rates and other observations
might be due to opacity of the cometary comae or may be due to assumed input parameters
 in the derivation of gas production rates.
Under assumed optically thin condition the \cite{Ootsubo12} derived gas production 
rates in several comets can be regarded as lower limits. Considering these
observational facts we varied CO$_2$ abundances in the model to assess our model
calculated \grat\ with the  observations. By increasing CO$_2$ abundances in these
comets by a factor of 3 we could achieve better  agreement with the observed  \grat.

Similarly, the calculations presented in Fig.~\ref{fig:gr-co2} demonstrate that  
for a constant H$_2$O production rate, the \grat\ increases with increasing CO$_2$
relative abundance. This figure suggests that for  a constant CO$_2$ relative
abundance, by increasing the H$_2$O production rate, the collisional quenching of
\ols\ and \old\ can increase the \grat.
Thus, the observation of a larger \grat\ value need not be  always due to higher 
CO$_2$ abundances.

In the case of hypothetical  comet X, which has CO$_2$ abundance equal to that of
H$_2$O,
the calculated percentage contributions of different processes to
red-doublet emissions  presented in Table~\ref{tab-slit} suggest that the red-doublet
 emission intensity is equally
controlled  by CO$_2$ and H$_2$O. If a comet has equal abundances of CO$_2$ 
and H$_2$O,  which is the case for comet \chrisw\ observed by \cite{Ootsubo12}  at
3.7 au from the Sun,
deriving the water production rates based on the observed red-doublet emission 
intensity may result in
over estimation of H$_2$O. In this case the derivation of CO$_2$ abundances using 
the observed \grat\ also leads to
improper  estimation.
This calculation suggests that in a comet having high CO$_2$ abundance, the
red-doublet emission intensity is not suitable for measuring H$_2$O rates.
Similarly, our model calculations on comet \neatq, which has 75\% CO$_2$ relative
abundance, suggest that around 30\% of
red-doublet emissions are governed by both photodissociation of CO$_2$ and radiative 
decay of \ols, which is comparable to the contribution from H$_2$O ($\sim$45\%)
(see Table~\ref{tab-slit}).

\subsection{Impact of collisional quenching of \ols\ and \old\ on the \grat}
The \grat\ at a given projected distance  mainly  depends on the formation
and destruction processes of excited oxygen atoms in the cometary coma
along the line of sight. The
abundances of O-bearing species and solar flux governs the formation rate of these 
metastable species
while the chemical lifetime and collisional quenching by other cometary species
determines the destruction rate. In a comet
having moderate H$_2$O production rate of 4 $\times$ 10$^{28}$ s$^{-1}$, the radius 
of the H$_2$O collisional zone is around 1000 km \citep{Whipple76}.
When the comet is at a larger heliocentric distance, a lower gas evaporation rate
results in a smaller collisional coma.
Discarding the collisional quenching effect
the observed  \grat\ has been used
to infer CO$_2$ production rate in comets observed at large heliocentric distances 
\citep{McKay12,Decock13}.
Our calculated \grat\ values as a function of projected distance on different comets
(cf. Figures~\ref{fig:rg-inten} and \ref{fig:grat-prj})
have shown that the collisional quenching of \ols\ and \old\ can result in larger 
(even $>$1)
 \grat\ values. 
Since the \grat\ is averaged over the observed large projected distances, the 
collisional quenching may not
influence the average value. In this case the observed \grat\ is mainly determined by
photochemical reactions of H$_2$O and CO$_2$ in producing red and green emissions, 
respectively.  Hence, the observed \grat\ value can be used to estimate the upper
limit of CO$_2$ abundance relative to the H$_2$O production rate. But in the case of
observations over  smaller projected distances the collisional
zone can predominantly  affect the observed \grat\ value which eventually can lead
to the estimation of higher CO$_2$ abundances. Since the comets considered in this
study are observed over large projected distances the effect of  collisional
quenching is
small on the averaged \grat. In such cases the observed \grat\  value can be
effective in constraining the upper limit of the CO$_2$ relative abundance.

\subsection{Green and red-doublet emission line widths}
 \cite{Cochran08} made high-resolution observations on different comets and found
that  the green line
 width is higher than both red-doublet emission lines. The observation of
these forbidden lines made on 12 comets have also shown the same feature 
\citep{Decock13}.
 The wider green line implies higher mean velocity distribution of O($^1$S)  atoms in
 the cometary coma. The high velocity of \ols\ atoms in the cometary coma  could be 
due to a 
parent source other than H$_2$O or could be due to involvement of high energy photons
in H$_2$O dissociation.
Our model calculations on comet \hbop\ showed that CO$_2$ photodissociation is a
potentially more
important source than that of H$_2$O in
 producing \ols\ atoms with high excess velocity \citep{Raghuram13}.

From the  calculations presented in Table~\ref{tab-slit},  it can be understood that
both CO$_2$ and H$_2$O are the important sources of \ols, whereas \old\ is mainly 
sourced from H$_2$O.
Since high energy photons (955--1165 \AA) mainly
dissociate CO$_2$ and produce \ols, the mean excess energy released in this reaction 
is larger
($\sim$2.5 eV) compared to that of H$_2$O ($\sim$1.2 eV). This results in the 
production of
 \ols\ atoms with large velocities (4.3 km s$^{-1}$) in cometary coma.
The calculations presented in Table~\ref{tab-slit} show that above 10$^4$ km 
projected distances,
 the thermal recombination of H$_2$O$^+$ and CO$_2^+$ ions together results in the 
production of
 15--40\% of \ols\ and around 20\% of \old.
\cite{Rosen00} and \cite{Seiersen03} experimentally determined the excess energies
and branching ratios for the dissociative products in dissociative recombination
of H$_2$O$^+$ and CO$_2^+$ ions, respectively.
 Based on these measured branching fractions and
by theoretical estimation, we calculated the excess velocities
of \ols\ and \old\ and green and red line widths by incorporating the dissociative 
recombination
reactions. We found an increase in our calculated green and red line widths by a factor
of 1.2--1.7 and 1.1--2.2, respectively. However, without accounting for dissociative 
recombination
reactions in our model, the calculated \grat\ values (see~Table~\ref{tab-comet})
 and line widths (see~Table~\ref{tab-slit}) are consistent with the observations.
In comet \neatq\ our calculated red line width is smaller than the observed value.
It can be noticed that in this comet both green and red-doublet
line widths are nearly the same and the red line widths are higher compared to
those on other comets, which is difficult to explain based on the model
calculations.

Our calculations show that the dissociative recombination of H$_2$O$^+$ and CO$_2^+$
ions are an important source of \ols\ in the outer coma. In the model calculations
 we assumed a constant electron-ion recombination temperature of 300 K. Since comets are
observed at large heliocentric distances the temperature values can be less than 300 K. 
To study the effect of electron temperature on the calculated \grat\ and line widths
we decreased the temperature to 200 K. We did not find any noticeable change in the 
calculated
 \grat\ values or line widths. Since most of the green and
red-doublet emission intensities are determined by photodissociation reactions in the
 inner
coma the contribution of thermal recombination of ions on the averaged \grat\ is
rather  small.

Several observations beyond 2 au have shown that the H$_2$O production rate in
comets
 does not vary as a function of the inverse square of
heliocentric distance \citep{Biver97,Biver99,Biver07,Bodewits12b}. Hence, extrapolation
of the H$_2$O production rate based on an approximation of the inverse square of
heliocentric 
distance 
may be inappropriate. We evaluated the implications of this extrapolation in comet 
\linek\
by decreasing the H$_2$O production rate by a factor of 2. No significant change 
(decrease by $\le$ 5\%)
is observed in the model calculated \grat\ and the calculated line widths.  

\subsection{Effect of atmospheric seeing}
For small (0.6--2$''$) slit observations the differential atmospheric seeing
can be an issue while determining the \grat\ based on the observed atomic oxygen
green and red-doublet line emission fluxes. The work carried out by \cite{McKay14}
suggests that the differential refraction is potentially important for the near UV
(e.g. CN 3870 \AA) compared to the observation in the wavelength region 5000--6500
\AA. They calculated the effect of differential refraction to be around 5\% or
less. In order to estimate the atmospheric seeing effect in determining slit-averaged
\grat\ we convolved our model calculated green and red-doublet emission fluxes with
Gaussian function with full width at half maximum of seeing value 1.0$''$. All these
comets, except \wildp, considered in the present work have been observed at
larger ($>$2 au) geocentric distances; hence, the projected area on these comets
would be larger.
 
\begin{center}
\begin{figure}
\noindent\includegraphics[width=22pc,angle=0]{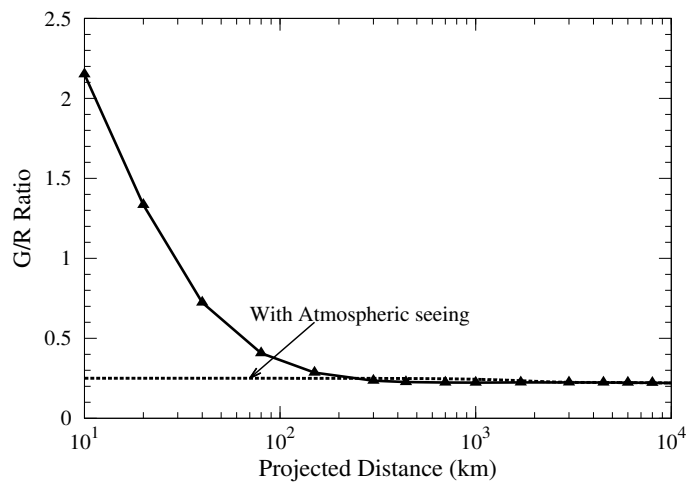}
\caption{The model calculated \grat\ on
comet \neat\ at r = 3.7 au and $\Delta$ = 3.4 au. The black dashed line is the
calculated \grat\ after convolving with a Gaussian function of full width at half
maximum of 1.0$''$ seeing value. }
\label{fig:gr-neat}
\end{figure}
\end{center}

In comet \neatq, which was observed at r = 3.7 au and $\Delta$ 
= 3.4 au, the model calculated G/R ratio varies between 2.2 and 0.2 below 100
km projected distance (cf. Figures \ref{fig:grat-prj} and \ref{fig:gr-neat}). After
incorporating the seeing effect we found that the calculated \grat\ is a 
constant throughout the projected distances with a value of about 0.2 as
shown in Fig.~\ref{fig:gr-neat}. However, we do not find any change in the
slit-averaged \grat\ after accounting for the seeing effect in the model. Since the 
slit-averaged \grat\ is over a much larger projected area, while the seeing effect
is confined to distances close to the nucleus (cf. Fig.~\ref{fig:grat-prj}). We also
assessed the output by changing the seeing value (from 1.0$''$ to 0.5$''$). No
appreciable change in the modelled \grat\ is observed. 
This suggests that the atmospheric seeing effect does not influence the \grat\ for
the
comets observed at larger ($>$2 au) geocentric distances. Detailed analysis of the
atmospheric seeing on different comets observed at different geocentric distances
($<$ 2 au) are being carried out and will be presented in our next paper (Decock et
al. 2014, in preparation).

\section{Summary and conclusion}
The observation of green and red-doublet emission lines in comets at larger ($>$ 2
au)  heliocentric distances suggest that the \grat\ value is larger than 0.1.
Moreover, the high-resolution observation  
reports that the green line is wider than the red-doublet lines, which
is difficult to explain based on the single parent source for these oxygen emission 
lines \citep{Decock13}.
 We have developed a coupled chemistry-emission model for atomic oxygen visible
prompt  emissions and applied it on six comets, (viz. \wild, \linear, \sid, \chris,
\garad, \neat)  which are observed at heliocentric distances greater than 2 au.
By accounting for important chemical reactions in the model we calculated the \grat\ 
values and widths of green and red-doublet emission lines on these comets.
It is found that CO$_2$ is potentially more important than H$_2$O in \ols\ production
 while \old\ is mainly controlled by H$_2$O. The photodissociation of CO is an
insignificant  source of metastable oxygen atoms. The observed large green line width
in several comets is due to higher velocity of \ols\ atoms that are essentially
produced via photodissociation of CO$_2$ by higher energy (955-1165 \AA) photons.
We have shown that the collisional quenching of \ols\ and \old\ by H$_2$O can lead 
to a larger \grat\ value and that its impact on the \grat\ is larger than the change
in CO$_2$ relative abundance.
Hence, the  larger \grat\ value need not always be linked to larger CO$_2$ abundances.
In a comet having large ($>$50\%) CO$_2$ abundances, the  photodissociation of CO$_2$
 plays a significant role
in producing both green and red-doublet emissions; thus, this process should also be 
accounted for while deriving the
H$_2$O production rate based on the red-doublet emission intensity. When a comet is 
observed over a larger projected distance where the collisional zone is less
resolvable, the  collisional quenching does not affect the observed \grat.  At larger
heliocentric distances, due to smaller gas production rates, the radius of a
collisional coma is smaller;  hence, the \grat\ observed over larger projected
distances  can be used to constrain the CO$_2$ relative abundance. However,  if the
slit-projected area on the comet is smaller  (with respect to the collisional zone),
the derived CO$_2$ abundance based on the \grat\ would be overestimated. Our model
calculated \grat\ and line widths of green
and red-doublet emission are in agreement  with the observation.

\begin{acknowledgements}
S. Raghuram was supported by the ISRO Senior Research Fellowship during the period
 of this work. Solar Irradiance Platform historical irradiances are provided courtesy
 of W.~Kent Tobiska and Space Environment Technologies. These historical irradiances
have been
developed with partial funding from the NASA UARS, TIMED, and SOHO missions. 
We thank Alice Decock, Jeffrey P. Morgenthaler, and Adam McKay for the
helpful discussions on the seeing effect on observation of comets.
\end{acknowledgements}

%   \bibliographystyle{aa}
%    \bibliography{/home/raghuram/Work/mypaper/bib-refferences/references-raghu-old}

\end{document}